\documentclass[aps,prb,twocolumn]{revtex4}
\usepackage{bm}
\usepackage{tabularx}
\usepackage{ascmac}
\usepackage{amsmath}
\usepackage{wrapfig}
\usepackage{graphicx}
\usepackage{float}
\usepackage[FIGBOTCAP]{subfigure}
\usepackage{color}

\bmdefine{\boldb}{b}
\bmdefine{\bolds}{s}
\bmdefine{\boldS}{S}
\bmdefine{\boldso}{so}
\bmdefine{\boldSO}{SO}
\bmdefine{\boldi}{i}
\bmdefine{\boldj}{j}
\bmdefine{\boldtau}{\tau}
\bmdefine{\boldsigma}{\sigma}
\bmdefine{\boldl}{l}
\bmdefine{\boldH}{H}
\bmdefine{\boldL}{L}
\bmdefine{\boldnabla}{\nabla}
\bmdefine{\boldlambda}{\lambda}
\bmdefine{\boldx}{x}
\bmdefine{\boldX}{X}
\bmdefine{\boldk}{k}
\bmdefine{\boldK}{K}
\bmdefine{\boldp}{p}
\bmdefine{\boldq}{q}
\bmdefine{\boldQ}{Q}
\bmdefine{\boldD}{D}
\bmdefine{\boldr}{r}
\bmdefine{\boldR}{R}
\bmdefine{\boldn}{n}
\bmdefine{\boldj}{j}
\bmdefine{\boldA}{A}
\bmdefine{\boldzero}{0}
\bmdefine{\boldone}{1}
\bmdefine{\boldtwo}{2}
\bmdefine{\boldthree}{3}
\bmdefine{\boldfour}{4}
\begin{document}

%\preprint{}

\title{
Tunneling between chiral magnets: Spin current generation
without external fields
}


\author{Naoya Arakawa}
\email{arakawa@hosi.phys.s.u-tokyo.ac.jp} 
\affiliation{
Center for Emergent Matter Science (CEMS), 
RIKEN, Wako, Saitama 351-0198, Japan}
%\thanks{}
%\altaffiliation{a}


\begin{abstract}
Magnons can generate a spin current, 
and the standard generating mechanism requires 
at least one external field. 
Since this mechanism is often applied to a multilayer system 
including a magnet and a paramagnetic metal, 
the system can possess not only the charge current induced by the spin current 
but also the charge current induced by the external field. 
The latter is an unnecessary accompaniment. 
Here we show that the tunneling of a magnon pair 
between chiral magnets can generate  
a spin current even without external fields. 
This phenomenon originates from 
a phase difference between magnon pairs of separate, weakly coupled chiral magnets, 
and is essentially different from the mechanism 
using the angle degree of freedom of the magnon Bose-Einstein condensates. 
The pair's tunneling is possible in chiral magnets 
due to lack of the Goldstone-type gapless excitations.  
This phenomenon opens the door to spintronics not requiring 
any external field and using the magnon pair tunneling.
\end{abstract}

%75.30.Ds: Spin waves
%75.76.+j: Spin transport effects
%75.30.-m: Intrinsic properties in magnetically ordered materials

\date{\today}
\maketitle



\section{Introduction}
Spin transport phenomena 
use a spin current~\cite{review-Spintronics}. 
The spin current is a flow of the spin angular momentum. 
Typically, 
its carriers for metals or semiconductors 
are conducting electrons, 
and the carriers for insulators are magnons; 
magnons are 
bosonic quasiparticles describing 
collective excitations in a magnet, 
a magnetically ordered insulator~\cite{Yosida-text,Kanamori-text}. 
The spin current for magnets can often flow over a larger distance 
than for metals or semiconductors~\cite{SpinDiffusion}. 
In addition, 
magnets have another advantage: 
charge transport is never accompanied 
due to lack of charge degrees of freedom. 
Because of those advantages, 
the spin transport phenomena using a magnet 
have been studied extensively~\cite{JS-InsFM1,JS-SSE,JS-InsFM2,JS-AF,Berger,Maekawa1,
Sigrist,Manske,MagBEC1,MagBEC2}. 

To generate the spin current in a magnet, 
we often apply an external field. 
For example, 
we consider a ferromagnet~\cite{JS-InsFM1,JS-InsFM2}. 
If an external magnetic field fulfills the resonance conditions~\cite{FMresonance}, 
it causes precession of the magnetization, 
and then this precession induces the spin current. 
This spin current is detectable, for example, 
by measuring the voltage due to the inverse spin Hall effect 
in a bilayer system of a ferromagnet and a metal~\cite{JS-InsFM1}. 
This is because the spin current in the ferromagnet layer 
pumps the spin current in the metal layer, 
and the latter can cause the charge current perpendicular. 
An external field is necessary even for an antiferromagnet~\cite{JS-AF}. 

The standard method of generating the spin current in a magnet 
possesses an unnecessary accompaniment. 
The standard method uses a multilayer system 
and at least one external field; 
the multilayer system includes at least one magnet layer and one metal layer; 
the external field is applied to the whole system. 
Thus, 
the external field can induce not only spin transport 
but also charge transport in the metal layer; 
this charge transport is distinct from 
the charge transport induced by the spin current. 
For example, 
in a bilayer system, 
the external magnetic field can induce 
the charge Hall effect 
in combination with the charge current induced by the spin current 
as long as 
that field is not parallel to the charge current.  
Furthermore, 
the accompanied charge transport may contribute to the final output. 
Actually, 
the voltage generation using a temperature gradient 
in a bilayer system includes both 
the contribution of the spin Seebeck effect in the magnet layer
and inverse spin Hall effect in the metal layer
and the contribution of the charge Seebeck effect in the metal layer~\cite{JS-SSE,JS-AF}. 
The electronic structure of the metal layer determines 
whether or not the latter is negligible, and 
whether the latter decreases or increases the total. 

A method not requiring external fields 
may be useful to utilize an advantage of magnets, 
lack of charge transport,  
as much as possible in a multilayer system. 
This is because 
such a method is free from charge transport accompanied by the external field. 

\begin{figure}[tb]
\includegraphics[width=60mm]{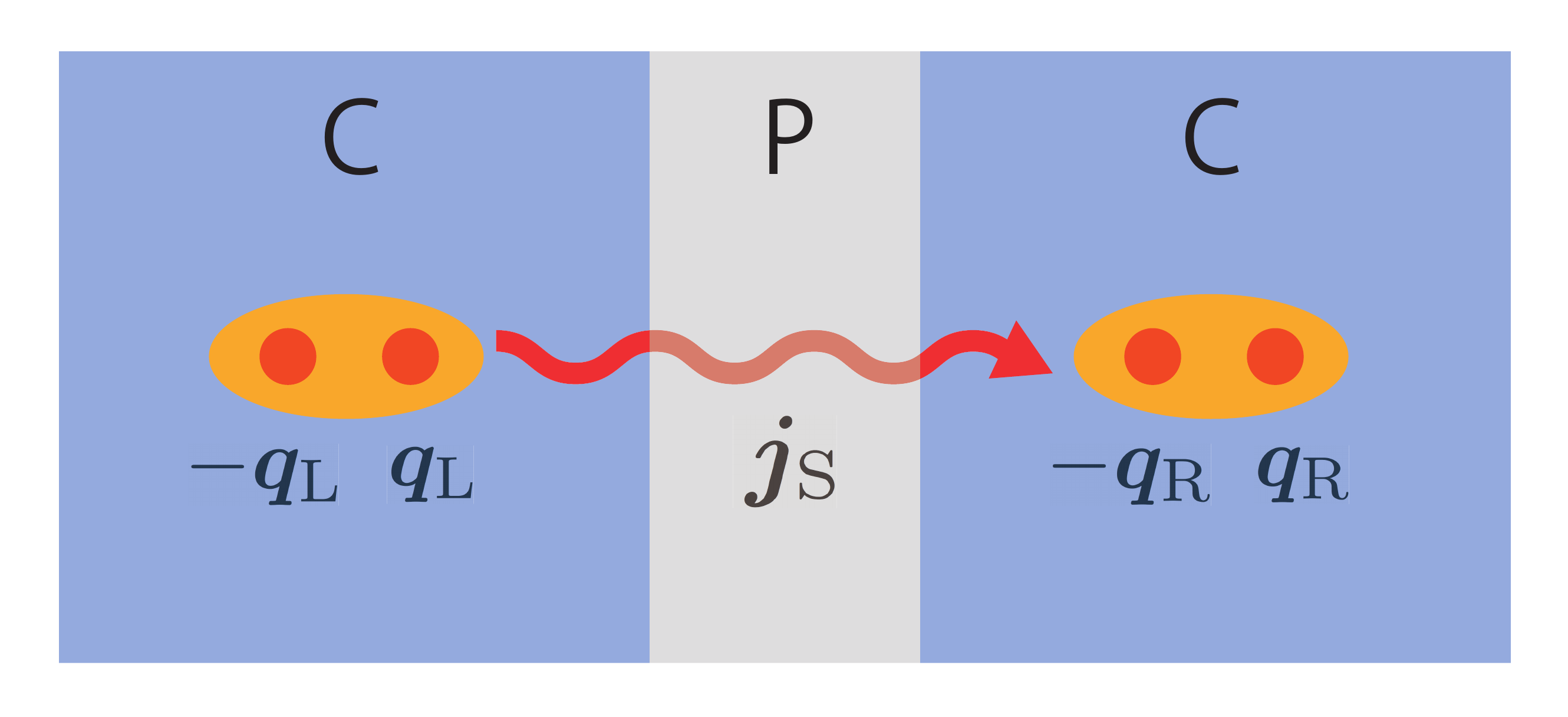}
\caption{
Schematic illustrations of our phenomenon. 
C denotes a chiral magnet, P denotes a paramagnetic metal or insulator or semiconductor, 
and $\boldj_{\textrm{S}}$ is the spin current due to the tunneling of a magnon pair. 
}
\label{fig1}
\end{figure}

Here we show that the tunneling between chiral magnets can generate 
the spin current without external fields (see Fig. \ref{fig1}). 
This originates from the tunneling of a magnon pair, 
which is present in chiral magnets but absent in ferromangets and antiferromagnets. 
This phenomenon is analogous to the Josephson effect~\cite{Joseph,Sigrist-Joseph}, 
which originates from the tunneling of a Cooper pair. 

\section{Basic concept}
\begin{figure}[tb]
\includegraphics[width=86mm]{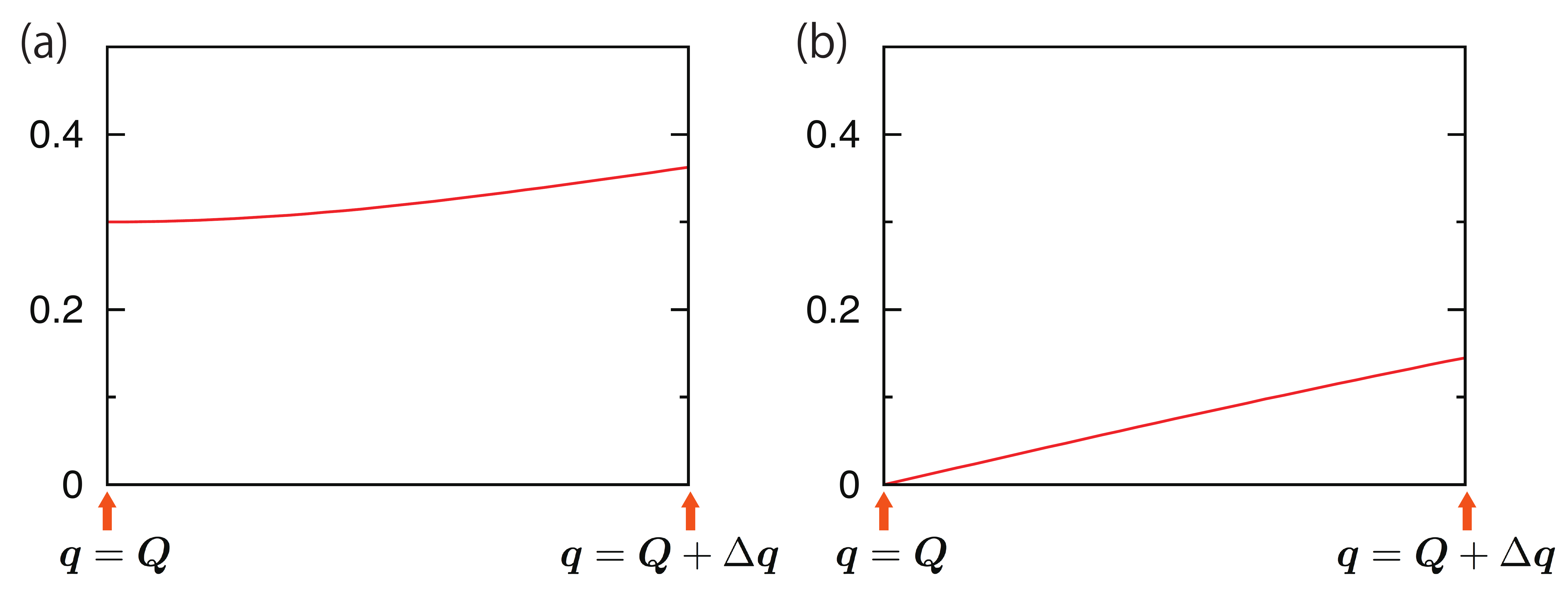}
\caption{
Schematic illustrations of the lowest-energy magnon branch around 
the ordering vector $\boldQ$ for 
(a) a chiral magnet and (b) a nonchiral magnet. 
The Goldstone type gapless excitation is absent in panel (a), 
and present in panel (b). 
}
\label{fig2}
\end{figure}

We start the basic concept of this phenomenon. 
The chiral magnet is a magnet having the spin scalar chirality, 
$\langle \hat{\boldS}_{\boldi}\cdot 
(\hat{\boldS}_{\boldj}\times \hat{\boldS}_{\boldk})\rangle$,  
where $\hat{\boldS}$ are spin operators, 
and $\boldi$, $\boldj$, $\boldk$ are site indices. 
This magnet also has finite three components of $\langle \hat{S}^{\alpha}_{\boldi}\rangle$
for $\alpha=x,y,z$, 
and the number of the finite components is invariant 
under global rotations in the spin space. 
This property results in a characteristic property of chiral magnets, 
the absence of the Goldstone-type gapless excitations of magnons~\cite{AIAO-exp,NA-LSWA-Pyro}, 
which holds even without magnetic anisotropy terms. 
This property contrasts with the presence in ferromagnets and antiferromagnets 
[compare Figs. \ref{fig2}(a) and \ref{fig2}(b)]; 
for those nonchiral magnets, 
the global rotation can change the number of the finite $\langle \hat{S}^{\alpha}_{\boldi}\rangle$. 
(Here we consider only the chiral or nonchiral magnet 
which is most stable without quantum fluctuations for a certain model, 
i.e., do not consider the magnets which are not most stable, 
because the stable solution of the magnon properties 
can be obtained only for the most stable state.)  
No Goldstone-type gapless excitations in the absence of magnetic anisotropy terms 
mean the finite expectation values of a magnon pair 
because the creation and annihilation operators of a magnon pair 
in the Hamiltonian act like the operators of the superconducting gap 
[compare Eqs. (\ref{eq:H_BCS}) and (\ref{eq:H_LSWA})]. 
The magnon pair is analogous to a Cooper pair in a superconductor~\cite{BCS}. 
This close relationship can be understood 
if we recall Anderson's pseudospin representation of the BCS Hamiltonian~\cite{Anderson-BCS}. 
The BCS Hamiltonian, a mean-field Hamiltonian for a superconductor, 
is expressed in terms of the pseudospins $\hat{\boldsigma}_{\boldk}$:
\begin{align}
\hat{H}_{\textrm{BCS}}
=&
\sum\limits_{\boldk,s}\xi_{\boldk}\hat{c}_{\boldk s}^{\dagger}\hat{c}_{\boldk s}
-\sum\limits_{\boldk}
(\Delta_{\boldk}\hat{c}_{-\boldk \downarrow}^{\dagger}\hat{c}_{\boldk\uparrow}^{\dagger}
+
\Delta_{\boldk}^{\ast}\hat{c}_{\boldk \uparrow}\hat{c}_{-\boldk \downarrow})\notag\\
=&
\sum\limits_{\boldk}
[
\xi_{\boldk}(1-\hat{\sigma}_{\boldk}^{z})
-2\textrm{Re}\Delta_{\boldk}\hat{\sigma}_{\boldk}^{x}
-2\textrm{Im}\Delta_{\boldk}\hat{\sigma}_{\boldk}^{y}].\label{eq:H_BCS}
\end{align}
Namely, a superconductor has not only the finite $z$ component of $\hat{\boldsigma}_{\boldk}$
but also the finite $x$ and $y$ components. 
(The absence of the Goldstone-type gapless excitation in superconductors 
can be understood as the invariance of the number of the finite $\hat{\boldsigma}_{\boldk}$ 
against global rotations.) 
The analogous property thus suggests the following prediction: 
\textit{the tunneling of a magnon pair can induce 
the spin current even without external fields.} 
This prediction is based on the analogy with 
the Josephson effect~\cite{Joseph}. 
This predicted phenomenon originates from 
neither  
the Bose-Einstein condensation (BEC) of magnons 
nor 
the effective magnetic field of $\langle \hat{\boldS}_{\boldi}\cdot 
(\hat{\boldS}_{\boldj}\times \hat{\boldS}_{\boldk})\rangle$. 
This can be understood if we recall two properties: 
the superconductivity cannot be regarded as the BEC 
of Cooper pairs due to the Pauli principle of electrons~\cite{Schrieffer-text}; 
the Josephson effect occurs even without any magnetic fields~\cite{SCtunnel-Ambeg}. 
Namely, 
the vital component in the Josephson effect is 
neither the BEC nor some magnetic field, 
but the finite probability of the pair's tunneling. 
Note, first, that 
the expectation value of a magnon pair can be finite
even without the magnon BEC (for more details, see Sec. IV); 
second, that a magnon pair and the spin scalar chirality 
result from exchange interactions stabilizing a chiral order,
i.e., either one of the two is not the cause for the other. 

\section{Derivation}
To demonstrate the phenomenon predicted above, 
we derive the spin current due to the tunneling of a magnon pair 
in a trilayer system. 
This system consists of two layers of a chiral magnet and 
one intermediate layer of a paramagnetic metal or insulator or semiconductor, 
as shown in Fig. \ref{fig1}. 

We describe the tunneling in this system 
using the following Hamiltonian:
$\hat{H}=\hat{H}_{\textrm{L}}+\hat{H}_{\textrm{R}}+\hat{V}$.
$\hat{H}_{\textrm{L}}$ and $\hat{H}_{\textrm{R}}$ 
are the Hamiltonians of a chiral magnet, 
and $\hat{V}$ is the tunneling Hamiltonian. 
For simplicity's sake, 
we choose the linearized-spin-wave Hamiltonians~\cite{Colpa,LSWA1,LSWA2,NA-LSWA-Pyro} as 
$\hat{H}_{\textrm{L}}$ and $\hat{H}_{\textrm{R}}$: 
$\hat{H}_{a}=\sum\textstyle_{\boldq_{a}}\hat{H}_{\textrm{LSW}}(\boldq_{a})$ 
($a=$L, R) with 
\begin{align}
\hat{H}_{\textrm{LSW}}(\boldq)
&=S\sum\limits_{l,l^{\prime}=1}^{N_{\textrm{sub}}}
[A_{ll^{\prime}}(\boldq)
\hat{b}_{\boldq l}^{\dagger}\hat{b}_{\boldq l^{\prime}}
+A_{ll^{\prime}}^{\ast}(-\boldq)
\hat{b}_{-\boldq l}\hat{b}_{-\boldq l^{\prime}}^{\dagger}]\notag\\
+&S\sum\limits_{l,l^{\prime}=1}^{N_{\textrm{sub}}}[
B_{ll^{\prime}}(\boldq)
\hat{b}_{\boldq l}^{\dagger}\hat{b}_{-\boldq l^{\prime}}^{\dagger}
+B_{ll^{\prime}}^{\ast}(-\boldq)
\hat{b}_{-\boldq l}\hat{b}_{\boldq l^{\prime}}].\label{eq:H_LSWA}
\end{align}
Here $\boldq_{\textrm{L}}$ and $\boldq_{\textrm{R}}$ 
are momenta in the left and right layers, 
$S$ is the total spin per site, 
$l$ and $l^{\prime}$ are sublattice indices, 
and $\hat{b}_{\boldq l}^{\dagger}$ and $\hat{b}_{\boldq^{\prime} l^{\prime}}$ 
are annihilation and creation operators of a magnon. 
If we compare Eqs. (\ref{eq:H_LSWA}) and (\ref{eq:H_BCS}), 
we can see that 
the terms of $B_{ll^{\prime}}(\boldq)$ and $B_{ll^{\prime}}^{\ast}(-\boldq)$ 
correspond to the terms of $\Delta_{\boldk}$ and $\Delta_{\boldk}^{\ast}$. 
We can express any exchange interactions of a magnet 
as Eq. (\ref{eq:H_LSWA}) 
by using the Holstein-Primakoff transformation 
if we neglect the terms of magnon-magnon interactions~\cite{LSWA2}. 
It is however necessary to keep in mind that 
Eq. (\ref{eq:H_LSWA}) should be derived for 
the most stable ordered state of a nonperturbative spin Hamiltonian 
without quantum fluctuations. 
Note that 
the linearized-spin-wave Hamiltonian for an antiferromagnet with two sublattices 
can be expressed without the $B_{ll^{\prime}}(\boldq)$ and $B_{ll^{\prime}}^{\ast}(-\boldq)$ terms 
if we set $\hat{b}_{\boldq l}=\hat{a}_{\boldq}$ for $l=A$ and $\hat{b}_{\boldq l}=\hat{b}_{\boldq}^{\dagger}$ 
for $l=B$. 
Then, 
we choose $\hat{V}$ as
\begin{align}
\hat{V}
=
2S
\sum\limits_{\boldq_{\textrm{L}},\boldq_{\textrm{R}}}
\sum\limits_{l=1}^{N_{\textrm{sub}}}
[
T_{\boldq_{\textrm{L}} \boldq_{\textrm{R}}}
\hat{b}_{\boldq_{\textrm{L}} l}^{\dagger}
\hat{b}_{\boldq_{\textrm{R}} l}
+
T_{\boldq_{\textrm{R}} \boldq_{\textrm{L}}}^{\ast}
\hat{b}_{\boldq_{\textrm{R}} l}^{\dagger}
\hat{b}_{\boldq_{\textrm{L}} l}
].\label{eq:H_T}
\end{align}
We assume that 
this term is finite and small, 
and then treat it as a second-order perturbation. 
This assumption physically means that 
the intermediate layer fulfills two conditions: 
its width is so thin that 
the overlap between the wave functions of a magnon 
in the left and right layers is finite; 
the width is so thick that 
the energy scale of the overlap 
is smaller than that of exchange interactions. 

The $\hat{V}$ changes the magnon numbers 
in the left and right layers, 
and then induces the spin current through the intermediate layer. 
The changes of the magnon numbers are determined 
by the Heisenberg equations of motion, 
$\dot{\hat{N}}_{\textrm{a}}=(i/\hbar)[\hat{V},\hat{N}_{\textrm{a}}]$ 
with $\hat{N}_{\textrm{a}}=\sum_{\boldq_{a},l} 
\hat{b}_{\boldq_{\textrm{a}} l}^{\dagger}\hat{b}_{\boldq_{\textrm{a}} l}$. 
The change for the right layer is given by
\begin{align}
&\langle \dot{\hat{N}}_{\textrm{R}}\rangle_{V} 
=-\frac{4S}{\hbar}
\sum\limits_{\boldq_{\textrm{L}},\boldq_{\textrm{R}}}
\sum\limits_{l=1}^{N_{\textrm{sub}}}
\textrm{Im}\bigl[
T_{\boldq_{\textrm{L}} \boldq_{\textrm{R}}}
\langle \hat{b}_{\boldq_{\textrm{L}} l}^{\dagger}(t)\hat{b}_{\boldq_{\textrm{R}} l}(t)\rangle_{V}
\bigr]\notag\\
&\approx \hspace{-3pt}
\frac{4S}{\hbar^{2}}
\hspace{-3pt}
\sum\limits
\hspace{-3pt}
\int_{-\infty}^{t}
\hspace{-12pt}
d t^{\prime}
e^{\alpha t^{\prime}}
\textrm{Re}
\bigl[T_{\boldq_{\textrm{L}} \boldq_{\textrm{R}}}
\langle [\hat{b}_{\boldq_{\textrm{L}} l}^{\dagger}(t)\hat{b}_{\boldq_{\textrm{R}} l}(t),
\hat{V}(t^{\prime})]\rangle
\bigr],\label{eq:NR-t}
\end{align}
where we have estimated $\langle \cdots \rangle_{V}$ 
within the first order of $\hat{V}$.  
After the integration, we take the limit $\alpha\rightarrow 0$. 
Then, the $\langle \dot{\hat{N}}_{\textrm{a}}\rangle_{V}$ 
is related to the spin current 
because the $\hat{b}_{\boldq_{\textrm{a}} l}^{\dagger}\hat{b}_{\boldq_{\textrm{a}} l}$ 
is expressed in terms of spin's $z$ component 
in the Holstein-Primakoff transformation~\cite{Colpa,LSWA1,LSWA2,NA-LSWA-Pyro}. 
We thus define the spin current as $J_{\textrm{S}}=\langle \dot{\hat{N}}_{\textrm{R}}\rangle_{V}$. 
This equation and Eq. (\ref{eq:NR-t}) 
suggest the $J_{\textrm{S}}$ can be induced by the $\hat{V}$.

Then, 
the $J_{\textrm{S}}$ can be divided up into two parts: 
$J_{\textrm{S}}=J_{\textrm{S}}^{(1)}+J_{\textrm{S}}^{(2)}$, 
where 
\begin{widetext}
\begin{align}
\hspace{-6pt}
J_{\textrm{S}}^{(1)}
\hspace{-2pt}
=&
\frac{8S^{2}}{\hbar^{2}}
\hspace{-4pt}
\sum\limits_{\boldq_{\textrm{L}},\boldq_{\textrm{R}}}
\sum\limits_{l,l^{\prime}}
|T_{\boldq_{\textrm{L}} \boldq_{\textrm{R}}}|^{2}
\hspace{-4pt}
\int_{-\infty}^{t}
\hspace{-10pt}
d t^{\prime}
e^{\delta t^{\prime}}
\textrm{Re}
\bigl[
\langle \hat{b}_{\boldq_{\textrm{L}} l}^{\dagger}(t)
\hat{b}_{\boldq_{\textrm{L}} l^{\prime}}(t^{\prime})\rangle
\langle \hat{b}_{\boldq_{\textrm{R}} l}(t)
\hat{b}_{\boldq_{\textrm{R}} l^{\prime}}^{\dagger}(t^{\prime})\rangle
-
\langle \hat{b}_{\boldq_{\textrm{R}} l^{\prime}}^{\dagger}(t^{\prime})
\hat{b}_{\boldq_{\textrm{R}} l}(t)\rangle
\langle \hat{b}_{\boldq_{\textrm{L}} l^{\prime}}(t^{\prime})
\hat{b}_{\boldq_{\textrm{L}} l}^{\dagger}(t)\rangle
\bigr],\label{eq:J_S^1}\\
\hspace{-6pt}
J_{\textrm{S}}^{(2)}
\hspace{-2pt}
=&
\frac{8S^{2}}{\hbar^{2}}
\hspace{-4pt}
\sum\limits_{\boldq_{\textrm{L}},\boldq_{\textrm{R}}}
\sum\limits_{l,l^{\prime}}
|T_{\boldq_{\textrm{L}} \boldq_{\textrm{R}}}|^{2}
\hspace{-4pt}
\int_{-\infty}^{t}
\hspace{-10pt}
d t^{\prime}
e^{\delta t^{\prime}}
\textrm{Re}
\bigl[
\langle \hat{b}_{\boldq_{\textrm{L}} l}^{\dagger}(t)
\hat{b}_{-\boldq_{\textrm{L}} l^{\prime}}^{\dagger}(t^{\prime})\rangle
\langle \hat{b}_{\boldq_{\textrm{R}} l}(t)
\hat{b}_{-\boldq_{\textrm{R}} l^{\prime}}(t^{\prime})\rangle
-
\langle \hat{b}_{-\boldq_{\textrm{L}} l^{\prime}}^{\dagger}(t^{\prime})
\hat{b}_{\boldq_{\textrm{L}} l}^{\dagger}(t)\rangle
\langle \hat{b}_{-\boldq_{\textrm{R}} l^{\prime}}(t^{\prime})
\hat{b}_{\boldq_{\textrm{R}} l}(t)\rangle
\bigr].\label{eq:J_S^2}
\end{align}
\end{widetext}
The $J_{\textrm{S}}^{(1)}$ and $J_{\textrm{S}}^{(2)}$ originate from, respectively, 
the tunneling of a single magnon 
and tunneling of a magnon pair. 
This is because the former includes $\langle \hat{b}^{\dagger}\hat{b}\rangle$ 
and $\langle \hat{b}\hat{b}^{\dagger}\rangle$, 
and the latter includes $\langle \hat{b}^{\dagger}\hat{b}^{\dagger}\rangle$ 
and $\langle \hat{b}\hat{b}\rangle$. 
Hereafter, 
we consider only the $J_{\textrm{S}}^{(2)}$ 
because the single-magnon's tunneling is possible even for nonchiral magnets, 
and the pair's tunneling is possible only for chiral magnets. 
Furthermore, 
the analogy with the Josephson effect~\cite{Joseph,SCtunnel-Ambeg,Sigrist-Joseph} suggests 
only the $J_{\textrm{S}}^{(2)}$ can be finite even without external fields.   

For further insight into the $J_{\textrm{S}}^{(2)}$, 
we rewrite Eq. (\ref{eq:J_S^2}) in a simpler form. 
In a similar way for the Josephson effect~\cite{SCtunnel-Ambeg}, 
we can express Eq. (\ref{eq:J_S^2}) in terms of 
the eigenvalues and eigenfunctions of $\hat{H}_{\textrm{L}}$ and $\hat{H}_{\textrm{R}}$ 
(for the details, see Appendix A):
\begin{align}
&J_{\textrm{S}}^{(2)}
=-\frac{16S^{2}}{\hbar}
\sum_{\boldq_{\textrm{L}},\boldq_{\textrm{R}}}
\sum_{l,l^{\prime}=1}^{N_{\textrm{sub}}}
\sum_{\nu,\nu^{\prime}=1}^{N_{\textrm{sub}}}
|T_{\boldq_{\textrm{L}} \boldq_{\textrm{R}}}|^{2}\notag\\
\times & P\Bigl(
\frac{n[\epsilon_{\nu}(\boldq_{\textrm{L}})]-n[\epsilon_{\nu^{\prime}}(\boldq_{\textrm{R}})]}
{\epsilon_{\nu}(\boldq_{\textrm{L}})-\epsilon_{\nu^{\prime}}(\boldq_{\textrm{R}})}
-
\frac{n[\epsilon_{\nu}(\boldq_{\textrm{L}})]-n[-\epsilon_{\nu^{\prime}}(\boldq_{\textrm{R}})]}
{\epsilon_{\nu}(\boldq_{\textrm{L}})+\epsilon_{\nu^{\prime}}(\boldq_{\textrm{R}})}
\Bigr)\notag\\
\times &
\textrm{Im}[
P_{\nu l}^{\dagger}(\boldq_{\textrm{L}})P_{l^{\prime}+N_{l} \nu}(\boldq_{\textrm{L}})
P_{l \nu^{\prime}}(\boldq_{\textrm{R}})P_{\nu^{\prime} l^{\prime}+N_{l}}^{\dagger}(\boldq_{\textrm{R}})
].\label{eq:J_S^2-simpler}
\end{align}
This is similar to the charge current~\cite{SCtunnel-Ambeg} due to the tunneling of a Cooper pair 
except for the differences in the constant coefficient and quasiparticles; 
the quasiparticles are fermions in superconductors. 
Due to the difference in quasiparticles, 
the spin current is zero at zero temperature. 
In Eq. (\ref{eq:J_S^2-simpler}), 
the eigenvalues, $\epsilon_{\nu}(\boldq)$, and eigenfunctions, $P_{\nu l}(\boldq)$, 
are given by
$\epsilon_{\nu}(\boldq)
=\sum_{l,l^{\prime}} 
[P^{\dagger}_{\nu l}(\boldq)A_{ll^{\prime}}(\boldq)P_{l^{\prime} \nu}(\boldq)
+P_{\nu l}^{\dagger}(\boldq)B_{ll^{\prime}}(\boldq)P_{l^{\prime}+N_{\textrm{sub}} \nu}(\boldq)
+P_{\nu l+N_{\textrm{sub}}}^{\dagger}(\boldq)B_{ll^{\prime}}^{\ast}(-\boldq)P_{l^{\prime} \nu}(\boldq)
+P_{\nu l+N_{\textrm{sub}}}^{\dagger}(\boldq)A_{ll^{\prime}}^{\ast}(-\boldq)P_{l^{\prime}+N_{\textrm{sub}} \nu}(\boldq)]$,
and $n(\epsilon)$ is the Bose-Einstein distribution function at temperature $T$,
$n(\epsilon)=(e^{\frac{\epsilon}{T}}-1)^{-1}$. 
Because of 
$P_{\nu l}^{\dagger}(\boldq_{\textrm{L}})P_{l^{\prime}+N_{\textrm{sub}} \nu}(\boldq_{\textrm{L}})
=\langle \nu|\hat{b}^{\dagger}_{\boldq_{\textrm{L}} l}\hat{b}^{\dagger}_{-\boldq_{\textrm{L}} l^{\prime}}|\nu \rangle
=C_{\boldq_{\textrm{L}}}^{\nu l l^{\prime}}e^{i\phi_{\boldq_{\textrm{L}}}^{\nu l l^{\prime}}}$ 
and 
$P_{l\nu^{\prime}}(\boldq_{\textrm{R}})P^{\dagger}_{\nu^{\prime}l^{\prime}+N_{\textrm{sub}}}(\boldq_{\textrm{R}})
=\langle \nu^{\prime}|\hat{b}_{-\boldq_{\textrm{R}} l}\hat{b}_{\boldq_{\textrm{R}} l^{\prime}}|\nu^{\prime}\rangle
=C_{\boldq_{\textrm{R}}}^{\nu^{\prime} ll^{\prime}}e^{-i\phi_{\boldq_{\textrm{R}}}^{\nu^{\prime}ll^{\prime}}}$, 
we can express Eq. (\ref{eq:J_S^2-simpler}) using 
the phase difference between 
the wave functions of a magnon pair in the left and right layers: 
\begin{align}
J_{\textrm{S}}^{(2)}
=
\sum_{\boldq_{\textrm{L}},\boldq_{\textrm{R}}}
\sum_{l,l^{\prime}=1}^{N_{\textrm{sub}}}
\sum_{\nu,\nu^{\prime}=1}^{N_{\textrm{sub}}}
J_{\boldq_{\textrm{L}} \boldq_{\textrm{R}}}^{ll^{\prime};\nu\nu^{\prime}}
\sin(\phi_{\boldq_{\textrm{L}}}^{\nu l l^{\prime}}-\phi_{\boldq_{\textrm{R}}}^{\nu^{\prime} l l^{\prime}}),
\label{eq:J_S^2-simpler2}
\end{align}
where 
$J_{\boldq_{\textrm{L}} \boldq_{\textrm{R}}}^{ll^{\prime};\nu\nu^{\prime}}
=-\frac{16S^{2}}{\hbar}
|T_{\boldq_{\textrm{L}} \boldq_{\textrm{R}}}|^{2}
P(
\frac{n[\epsilon_{\nu}(\boldq_{\textrm{L}})]-n[\epsilon_{\nu^{\prime}}(\boldq_{\textrm{R}})]}
{\epsilon_{\nu}(\boldq_{\textrm{L}})-\epsilon_{\nu^{\prime}}(\boldq_{\textrm{R}})}
-
\frac{n[\epsilon_{\nu}(\boldq_{\textrm{L}})]-n[-\epsilon_{\nu^{\prime}}(\boldq_{\textrm{R}})]}
{\epsilon_{\nu}(\boldq_{\textrm{L}})+\epsilon_{\nu^{\prime}}(\boldq_{\textrm{R}})}
)C_{\boldq_{\textrm{L}}}^{\nu ll^{\prime}}
C_{\boldq_{\textrm{R}}}^{\nu^{\prime} l l^{\prime}}$.
This phase difference can induce the spin current without external fields. 

\section{Discussion}
First, 
we discuss the validity of our theory. 
We used the linearized-spin-wave approximation for a general Hamiltonian 
of exchange interactions to treat magnons in a chiral magnet. 
This is valid 
if the ground state under the hard-spin constraint is nondegenerate, 
and if the temperature is so low that 
the interaction terms, neglected in this approximation, 
are negligible~\cite{LSWA2,NA-LSWA-Pyro}. 
The former condition is necessary 
because this approximation treats 
collective motions of spins as fluctuations 
against a most stable ground state under the hard-spin constraint~\cite{LSWA2,NA-LSWA-Pyro}. 
The latter is necessary because 
this approximation is a low-order expansion of the spin operators 
in terms of $\hat{b}^{\dagger}\hat{b}/2S$, 
which is small at low temperature~\cite{LSWA2,NA-LSWA-Pyro}. 
We also used Eq. (\ref{eq:H_T}) and its second-order perturbation 
to treat the tunneling between chiral magnets. 
This is valid 
if the left and right layers are weakly coupled 
due to the overlap between the wave functions of a magnon. 
Such weak coupling is probably reliable in experiments. 
Thus, our theory is valid for analyzing the tunneling between 
chiral magnets coupled weakly at low temperature. 

Next, 
we argue that 
our mechanism is distinct from the mechanism for generating 
the spin current in the magnon BEC 
(e.g., Refs.\onlinecite{Sonin,Sigrist,Manske,MagBEC1,MagBEC2,MacDonald,Tserkovnyak}). 
In the magnon BEC, 
$\langle \hat{b}_{\boldq}\rangle =O(N^{\frac{1}{2}})$ is fulfilled 
due to coherence of the wave function. 
Such coherence is preserved even in the presence of the magnon-magnon interactions 
as long as the interaction-induced damping is negligible compared with temperature. 
To generate the spin current in the magnon BEC, 
the angle degree of freedom of $\langle \hat{b}_{\boldj}\rangle$ 
(or $\langle \hat{b}^{\dagger}_{\boldj}\rangle$), 
$\theta_{\boldj}$ of $\langle \hat{b}_{\boldj}\rangle=B_{\boldj} e^{-i\theta_{\boldj}}$, is used; 
the spin current is proportional to $\boldnabla \theta_{\boldj}$.  
(For its example, see Appendix B.) 
While $\langle \hat{b}\hat{b}\rangle
=\langle \hat{b}\rangle \langle\hat{b}\rangle$ for noninteracting magnons, 
$\langle \hat{b}\hat{b}\rangle\neq 
\langle \hat{b}\rangle \langle\hat{b}\rangle$ for interacting magnons. 
In particular, 
$\langle \hat{b}\hat{b}\rangle$ and $\langle \hat{b}^{\dagger}\hat{b}^{\dagger}\rangle$ 
can be finite even in the absence of the magnon BEC. 
Namely, 
our mechanism can generate the spin current 
not only at low temperature, 
where the magnon BEC is present, 
but also at high temperature, 
where the magnon BEC is absent. 
Thus, our mechanism and the mechanism using the magnon BEC are essentially different. 

Then, 
we argue a method of testing our phenomenon by experiment. 
In our phenomenon, 
the spin current is generated in a multilayer system 
even without external fields. 
We can observe this spin current, 
for example, by using the inverse spin Hall effect in a paramagnetic metal 
of the intermediate layer 
in a similar way to Ref. \onlinecite{JS-InsFM1}. 
This is 
because near the interface between C and P in Fig. \ref{fig1} 
part of the magnon spin current can be converted into the electron spin current 
using the interfacial exchange interactions, 
and because chiral magnets can possess the non-degenerate magnon energy dispersion 
(i.e., we can avoid vanishing of the spin current due to 
the degeneracy of the magnon energies~\cite{Ohnuma}). 
However, there are several main differences 
between this and the standard phenomena~\cite{JS-InsFM1,JS-SSE,JS-InsFM2,JS-AF}; 
our phenomenon neither needs any external field 
nor has a charge current induced by it. 
As chiral magnets, 
we can use, for example, 
all-in/all-out, two-in-two-out, and three-in-one-out chiral magnets 
in pyrochlore oxides~\cite{NA-LSWA-Pyro}. 
While controlling the phase difference in our phenomenon 
is more difficult than controlling an external field, 
an analogy with the Josephson effect suggests that 
the phase difference can be finite using chiral magnets 
whose gap structures have different momentum dependence. 
(For the Josephson effect, 
one of the simple ways of obtaining the phase difference 
is to use superconductors whose gap structures are different, e.g., 
$s$-wave and $d$-wave superconductors.) 
Thus, our phenomenon is experimentally testable. 

Finally, 
we discuss implications of our phenomenon. 
Our phenomenon provides a method of generating the spin current 
in an insulator.  
Moreover, 
we can extend the formulation to 
study the high-temperature properties. 
At high temperature, 
we need to consider the magnon-magnon interactions, which 
cause spin-Coulomb drag~\cite{SCD-review,SCD-NA}, 
a characteristic dissipation of a nonconserved quantity. 
Thus, our phenomenon provides an opportunity of studying 
the spin-Coulomb drag. 
Then, 
a similar phenomenon is possible even for chiral metals, 
magnetically ordered metals with the spin scalar chirality: 
the tunneling between chiral metals can generate the spin current 
without external fields. 
There is however a significant difference: 
chiral metals have spin and charge degrees of freedom. 
It is thus desirable to reveal roles of charge degrees of freedom 
in the tunneling between chiral metals. 
In addition, 
knowledge of the Josephson effect implies some research. 
It is known that 
the Josephson effect in the presence of external dc and ac voltages 
is useful to determine the value of $e/\hbar$~\cite{Shapiro}. 
Also, it is known that 
the Josephson effect in the presence of an external magnetic field 
is useful to distinguish a $s$-wave gap and a $d$-wave gap~\cite{Joseph-dwave}. 
Thus, our phenomenon in the presence of some external field 
may be useful to determine the value of $1/\hbar$ (not including $e$) 
and the differences between the gap structures of different chiral magnets. 
Moreover, 
our phenomenon implies a similar phenomenon for phonons or photons. 
For example, 
the tunneling of a phonon pair may generate 
a charge current even without external fields. 

\section{Summary}
In summary, 
we have proposed that the tunneling of a magnon pair between chiral magnets 
can generate the spin current
even without external fields. 
We presented this proposal by deriving the spin current tunneling through 
the intermediate layer of the trilayer system of Fig. \ref{fig1}. 
In this derivation, 
we treated magnons in a chiral magnet using the linearized-spin-wave approximation. 
We also treated the weak tunneling between chiral magnets 
using the second-order perturbation. 
Our treatments are valid for the tunneling between weakly coupled chiral magnets 
at low temperature. 
Our phenomenon opens the door to 
spintronics not requiring any external field 
and using the magnon pair tunneling.


\appendix

\section{Derivation of Eq. (\ref{eq:J_S^2-simpler})}
We explain the detail of the derivation of Eq. (\ref{eq:J_S^2-simpler}). 

We first explain a method of obtaining the eigenvalues and eigenfunctions 
of $\hat{H}_{\textrm{L}}$ and $\hat{H}_{\textrm{R}}$. 
Since its detailed explanations have been provided, for example, 
in Refs. \onlinecite{Colpa} and \onlinecite{NA-LSWA-Pyro}, 
we here provide the brief explanation. 
We can diagonalize the Hamiltonian of a chiral magnet 
in the linearized-spin-wave approximation as follows:
\begin{widetext}
\begin{align}
\hat{H}_{a}
=&
S\sum\limits_{\boldq_{a}}
\sum\limits_{l,l^{\prime}=1}^{N_{\textrm{sub}}}
(\hat{b}_{\boldq_{a} l}^{\dagger}\ \hat{b}_{-\boldq_{a} l})
\left(
\begin{array}{@{\,}cc@{\,}}
A_{ll^{\prime}}(\boldq_{a}) & B_{ll^{\prime}}(\boldq_{a})\\
B^{\ast}_{ll^{\prime}}(-\boldq_{a}) & A^{\ast}_{ll^{\prime}}(-\boldq_{a})
\end{array} 
\right)
\left(
\begin{array}{@{\,}c@{\,}}
\hat{b}_{\boldq_{a} l^{\prime}}\\
\hat{b}_{-\boldq_{a} l^{\prime}}^{\dagger}
\end{array} 
\right)\notag\\
=&
S\sum\limits_{\boldq_{a}}
\sum\limits_{\nu=1}^{N_{\textrm{sub}}}
(\hat{b}_{\boldq_{a} \nu}^{\prime \dagger}\ \hat{b}_{-\boldq_{a} \nu+N_{\textrm{sub}}}^{\prime})
\left(
\begin{array}{@{\,}cc@{\,}}
\epsilon_{\nu}(\boldq_{a}) & 0\\
0 & \epsilon_{\nu+N_{\textrm{sub}}}(-\boldq_{a})
\end{array} 
\right)
\left(
\begin{array}{@{\,}c@{\,}}
\hat{b}_{\boldq_{a} \nu}^{\prime}\\
\hat{b}_{-\boldq_{a} \nu+N_{\textrm{sub}}}^{\prime\dagger}
\end{array} 
\right)
.\label{eq:HLSW}
\end{align}
\end{widetext}
Here $\hat{b}$ and $\hat{b}^{\prime}$ 
are connected in the following equation:
\begin{align}
\left(
\begin{array}{@{\,}c@{\,}}
\hat{b}_{\boldq l}\\
\hat{b}_{-\boldq l}^{\dagger}
\end{array} 
\right)
=
\sum\limits_{\nu=1}^{N_{\textrm{sub}}}
\left(
\begin{array}{@{\,}cc@{\,}}
P_{l\nu}(\boldq) & P_{l\nu+N_{\textrm{sub}}}(\boldq)\\
P_{l+N_{\textrm{sub}}\nu}(\boldq) & P_{l+N_{\textrm{sub}} \nu+N_{\textrm{sub}}}(\boldq)
\end{array} 
\right)
\left(
\begin{array}{@{\,}c@{\,}}
\hat{b}_{\boldq \nu}^{\prime}\\
\hat{b}_{-\boldq \nu+N_{\textrm{sub}}}^{\prime\dagger}
\end{array} 
\right),\label{eq:relation-b-b'}
\end{align}
where $P$ is a $(2N_{\textrm{sub}}\times 2N_{\textrm{sub}})$ paraunitary matrix. 
Note that 
paraunitary matrices should fulfill 
$PgP^{\dagger}=g$, 
where $g$ is the $(2N_{\textrm{sub}}\times 2N_{\textrm{sub}})$ paraunit matrix. 
Due to this property, 
the $P$ fulfills 
\begin{align}
\delta_{l,l^{\prime}}
&=\sum\limits_{\nu=1}^{N_{\textrm{sub}}}P_{l\nu}P_{\nu l^{\prime}}^{\dagger}
-\sum\limits_{\nu=1}^{N_{\textrm{sub}}}P_{l\nu+N_{\textrm{sub}}}P_{\nu+N_{\textrm{sub}} l^{\prime}}^{\dagger},
\label{eq:Psym1}\\
0
&=\sum\limits_{\nu=1}^{N_{\textrm{sub}}}P_{l\nu}P_{\nu l^{\prime}+N_{\textrm{sub}}}^{\dagger}
-\sum\limits_{\nu=1}^{N_{\textrm{sub}}}P_{l \nu+N_{\textrm{sub}}}P_{\nu+N_{\textrm{sub}} l^{\prime}+N_{\textrm{sub}}}^{\dagger},
\label{eq:Psym2}\\
0
&=\sum\limits_{\nu=1}^{N_{\textrm{sub}}}P_{l+N_{\textrm{sub}}\nu}P_{\nu l^{\prime}}^{\dagger}
-\sum\limits_{\nu=1}^{N_{\textrm{sub}}}P_{l+N_{\textrm{sub}} \nu+N_{\textrm{sub}}}P_{\nu+N_{\textrm{sub}} l^{\prime}}^{\dagger},
\label{eq:Psym3}\\
-\delta_{l,l^{\prime}}
&=\sum\limits_{\nu=1}^{N_{\textrm{sub}}}P_{l+N_{\textrm{sub}} \nu}P_{\nu l^{\prime}+N_{\textrm{sub}}}^{\dagger}
-\sum\limits_{\nu=1}^{N_{\textrm{sub}}}P_{l+N_{\textrm{sub}} \nu+N_{\textrm{sub}}}P_{\nu+N_{\textrm{sub}} l^{\prime}+N_{\textrm{sub}}}^{\dagger}
\label{eq:Psym4}.
\end{align} 
Combining Eqs. (\ref{eq:HLSW}) and (\ref{eq:relation-b-b'}), 
we obtain the eigenvalues, $\epsilon_{\nu}(\boldq_{a})$, 
and the eigenfunctions, $P_{l\nu}(\boldq_{a})$: 
\begin{align}
\epsilon_{\nu}(\boldq)
&=
\sum\limits_{l,l^{\prime}=1}^{N_{\textrm{sub}}} 
P^{\dagger}_{\nu l}(\boldq)A_{ll^{\prime}}(\boldq)P_{l^{\prime} \nu}(\boldq)\notag\\
&+\sum\limits_{l,l^{\prime}=1}^{N_{\textrm{sub}}} 
P_{\nu l}^{\dagger}(\boldq)B_{ll^{\prime}}(\boldq)P_{l^{\prime}+N_{\textrm{sub}} \nu}(\boldq)\notag\\
&+\sum\limits_{l,l^{\prime}=1}^{N_{\textrm{sub}}} 
P_{\nu l+N_{\textrm{sub}}}^{\dagger}(\boldq)B_{ll^{\prime}}^{\ast}(-\boldq)P_{l^{\prime} \nu}(\boldq)\notag\\
&+\sum\limits_{l,l^{\prime}=1}^{N_{\textrm{sub}}} 
P_{\nu l+N_{\textrm{sub}}}^{\dagger}(\boldq)A_{ll^{\prime}}^{\ast}(-\boldq)P_{l^{\prime}+N_{\textrm{sub}} \nu}(\boldq)
\label{eq:ev-ef}.
\end{align}
Note that $\epsilon_{\nu+N_{\textrm{sub}}}(-\boldq)$ 
is given by $\epsilon_{\nu+N_{\textrm{sub}}}(-\boldq)=\epsilon_{\nu}(\boldq)$.

Then, 
we rewrite Eq. (\ref{eq:J_S^2}) by using the eigenvalues and eigenfunctions 
of $\hat{H}_{\textrm{L}}$ and $\hat{H}_{\textrm{R}}$. 
For that purpose, we take five steps. 
First, we express the magnon operators in Eq. (\ref{eq:J_S^2}) 
in terms of $\hat{b}^{\prime}$ and $\hat{b}^{\prime \dagger}$. 
We obtain the expressions by using the relations, 
\begin{widetext}
\begin{align}
&\hat{b}_{\boldq l}(t)
=\sum\limits_{\nu=1}^{N_{\textrm{sub}}}
P_{l\nu}(\boldq)\hat{b}^{\prime}_{\boldq \nu}
e^{-\frac{i}{\hbar}\epsilon_{\nu}(\boldq)t}
+\sum\limits_{\nu=1}^{N_{\textrm{sub}}}
P_{l \nu+N_{\textrm{sub}}}(\boldq)\hat{b}_{-\boldq \nu+N_{\textrm{sub}}}^{\prime \dagger}
e^{\frac{i}{\hbar}\epsilon_{\nu}(\boldq)t},
\label{eq:b(t)1}\\
&\hat{b}_{-\boldq l}^{\dagger}(t)
=\sum\limits_{\nu=1}^{N_{\textrm{sub}}}
P_{l+N_{\textrm{sub}} \nu}(\boldq)\hat{b}_{\boldq \nu}^{\prime}
e^{-\frac{i}{\hbar}\epsilon_{\nu}(\boldq)t}
+\sum\limits_{\nu=1}^{N_{\textrm{sub}}}
P_{l+N_{\textrm{sub}} \nu+N_{\textrm{sub}}}(\boldq)\hat{b}_{-\boldq \nu+N_{\textrm{sub}}}^{\prime \dagger}
e^{\frac{i}{\hbar}\epsilon_{\nu}(\boldq)t}\label{eq:b(t)2}.
\end{align}
\end{widetext}
Namely, by using the above equations, we obtain
\begin{widetext}
\begin{align}
\langle \hat{b}_{\boldq_{\textrm{L}} l}^{\dagger}(t)
\hat{b}_{-\boldq_{\textrm{L}} l^{\prime}}^{\dagger}(t^{\prime}) \rangle
=&
\sum\limits_{\nu=1}^{N_{\textrm{sub}}}
P_{\nu l}^{\dagger}(\boldq_{\textrm{L}})P_{l^{\prime}+N_{\textrm{sub}} \nu}(\boldq_{\textrm{L}})
n[\epsilon_{\nu}(\boldq_{\textrm{L}})]e^{\frac{i}{\hbar}\epsilon_{\nu}(\boldq_{\textrm{L}})(t-t^{\prime})}\notag\\
&+
\sum\limits_{\nu=1}^{N_{\textrm{sub}}}
P_{\nu+N_{\textrm{sub}} l}^{\dagger}(\boldq_{\textrm{L}})
P_{l^{\prime}+N_{\textrm{sub}}\nu+N_{\textrm{sub}}}(\boldq_{\textrm{L}})
(1+n[\epsilon_{\nu}(\boldq_{\textrm{L}})])
e^{-\frac{i}{\hbar}\epsilon_{\nu}(\boldq_{\textrm{L}})(t-t^{\prime})},\label{eq:bb1}\\
\langle \hat{b}_{\boldq_{\textrm{R}} l}(t)
\hat{b}_{-\boldq_{\textrm{R}} l^{\prime}}(t^{\prime}) \rangle
=&
\sum\limits_{\nu=1}^{N_{\textrm{sub}}}
P^{\dagger}_{\nu l^{\prime}+N_{\textrm{sub}}}(\boldq_{\textrm{R}})P_{l\nu}(\boldq_{\textrm{R}})
(1+n[\epsilon_{\nu}(\boldq_{\textrm{R}})])e^{-\frac{i}{\hbar}\epsilon_{\nu}(\boldq_{\textrm{R}})(t-t^{\prime})}\notag\\
&+
\sum\limits_{\nu=1}^{N_{\textrm{sub}}}
P^{\dagger}_{\nu+N_{\textrm{sub}} l^{\prime}+N_{\textrm{sub}}}(\boldq_{\textrm{R}})
P_{l \nu+N_{\textrm{sub}}}(\boldq_{\textrm{R}})
n[\epsilon_{\nu}(\boldq_{\textrm{R}})]
e^{\frac{i}{\hbar}\epsilon_{\nu}(\boldq_{\textrm{R}})(t-t^{\prime})},\label{eq:bb2}\\
\langle \hat{b}_{-\boldq_{\textrm{R}} l^{\prime}}(t^{\prime})
\hat{b}_{\boldq_{\textrm{R}} l}(t) \rangle
=&
\sum\limits_{\nu=1}^{N_{\textrm{sub}}}
P^{\dagger}_{\nu l^{\prime}+N_{\textrm{sub}}}(\boldq_{\textrm{R}})
P_{l\nu}(\boldq_{\textrm{R}})
n[\epsilon_{\nu}(\boldq_{\textrm{R}})]e^{-\frac{i}{\hbar}\epsilon_{\nu}(\boldq_{\textrm{R}})(t-t^{\prime})}\notag\\
&+
\sum\limits_{\nu=1}^{N_{\textrm{sub}}}
P^{\dagger}_{\nu+N_{\textrm{sub}} l^{\prime}+N_{\textrm{sub}}}(\boldq_{\textrm{R}})
P_{l \nu+N_{\textrm{sub}}}(\boldq_{\textrm{R}})
(1+n[\epsilon_{\nu}(\boldq_{\textrm{R}})])
e^{\frac{i}{\hbar}\epsilon_{\nu}(\boldq_{\textrm{R}})(t-t^{\prime})},\label{eq:bb3}\\
\langle \hat{b}_{-\boldq_{\textrm{L}} l^{\prime}}^{\dagger}(t^{\prime})
\hat{b}_{\boldq_{\textrm{L}} l}^{\dagger}(t) \rangle
=&
\sum\limits_{\nu=1}^{N_{\textrm{sub}}}
P_{\nu l}^{\dagger}(\boldq_{\textrm{L}})P_{l^{\prime}+N_{\textrm{sub}} \nu}(\boldq_{\textrm{L}})
(1+n[\epsilon_{\nu}(\boldq_{\textrm{L}})])e^{\frac{i}{\hbar}\epsilon_{\nu}(\boldq_{\textrm{L}})(t-t^{\prime})}\notag\\
&+
\sum\limits_{\nu=1}^{N_{\textrm{sub}}}
P_{\nu+N_{\textrm{sub}} l}^{\dagger}(\boldq_{\textrm{L}})
P_{l^{\prime}+N_{\textrm{sub}}\nu+N_{\textrm{sub}}}(\boldq_{\textrm{L}})
n[\epsilon_{\nu}(\boldq_{\textrm{L}})]
e^{-\frac{i}{\hbar}\epsilon_{\nu}(\boldq_{\textrm{L}})(t-t^{\prime})}.\label{eq:bb4} 
\end{align}
\end{widetext}
Here we have used the relations, 
such as $\langle \hat{b}^{\prime \dagger}_{\boldq \nu}\hat{b}^{\prime}_{\boldq^{\prime} \nu^{\prime}}\rangle
=\delta_{\boldq,\boldq^{\prime}}\delta_{\nu,\nu^{\prime}}n[\epsilon_{\nu}(\boldq)]$. 
Second, 
we perform the integration about $t^{\prime}$ in Eq. (\ref{eq:J_S^2}). 
For the integration,  
it is sufficient to use the following results:
\begin{widetext}
\begin{align}
\int_{-\infty}^{t}
d t^{\prime}
e^{\alpha t^{\prime}}
e^{i[\epsilon_{\nu}(\boldq_{\textrm{L}})/\hbar-\epsilon_{\nu^{\prime}}(\boldq_{\textrm{R}})/\hbar](t-t^{\prime})}
&=iP\Bigl(
\frac{\hbar}{\epsilon_{\nu}(\boldq_{\textrm{L}})-\epsilon_{\nu^{\prime}}(\boldq_{\textrm{R}})}
\Bigr)
+\pi \delta
\Bigl(
\frac{\epsilon_{\nu}(\boldq_{\textrm{L}})-\epsilon_{\nu^{\prime}}(\boldq_{\textrm{R}})}{\hbar}
\Bigr),
\label{eq:int1}\\
-\int_{-\infty}^{t}
d t^{\prime}
e^{\alpha t^{\prime}}
e^{-i[\epsilon_{\nu}(\boldq_{\textrm{L}})/\hbar-\epsilon_{\nu^{\prime}}(\boldq_{\textrm{R}})/\hbar](t-t^{\prime})}
&=iP
\Bigl(
\frac{\hbar}{\epsilon_{\nu}(\boldq_{\textrm{L}})-\epsilon_{\nu^{\prime}}(\boldq_{\textrm{R}})}
\Bigr)
-\pi \delta
\Bigl(
\frac{\epsilon_{\nu}(\boldq_{\textrm{L}})-\epsilon_{\nu^{\prime}}(\boldq_{\textrm{R}})}{\hbar}
\Bigr),
\label{eq:int2}\\
-\int_{-\infty}^{t}
d t^{\prime}
e^{\alpha t^{\prime}}
e^{i[\epsilon_{\nu}(\boldq_{\textrm{L}})/\hbar+\epsilon_{\nu^{\prime}}(\boldq_{\textrm{R}})/\hbar](t-t^{\prime})}
&=-iP
\Bigl(
\frac{\hbar}{\epsilon_{\nu}(\boldq_{\textrm{L}})+\epsilon_{\nu^{\prime}}(\boldq_{\textrm{R}})}
\Bigr)
-\pi \delta
\Bigl(
\frac{\epsilon_{\nu}(\boldq_{\textrm{L}})+\epsilon_{\nu^{\prime}}(\boldq_{\textrm{R}})}{\hbar}
\Bigr),
\label{eq:int3}\\
\int_{-\infty}^{t}
d t^{\prime}
e^{\alpha t^{\prime}}
e^{-i[\epsilon_{\nu}(\boldq_{\textrm{L}})/\hbar+\epsilon_{\nu^{\prime}}(\boldq_{\textrm{R}})/\hbar](t-t^{\prime})}
&=-iP
\Bigl(
\frac{\hbar}{\epsilon_{\nu}(\boldq_{\textrm{L}})+\epsilon_{\nu^{\prime}}(\boldq_{\textrm{R}})}
\Bigr)
+\pi \delta
\Bigl(
\frac{\epsilon_{\nu}(\boldq_{\textrm{L}})+\epsilon_{\nu^{\prime}}(\boldq_{\textrm{R}})}{\hbar}
\Bigr).
\label{eq:int4}
\end{align}
\end{widetext}
After the integration, we took the limit $\alpha\rightarrow 0$. 
Third, we combine Eq. (\ref{eq:J_S^2}) with 
Eqs. (\ref{eq:bb1}){--}(\ref{eq:bb4}) and (\ref{eq:int1}){--}(\ref{eq:int4}). 
As the result, 
Eq. (\ref{eq:J_S^2}) becomes
\begin{widetext}
\begin{align}
J_{\textrm{S}}^{(2)}
&=
-\frac{8S^{2}}{\hbar}
\sum\limits_{\boldq_{\textrm{L}},\boldq_{\textrm{R}}}
\sum\limits_{l,l^{\prime}=1}^{N_{\textrm{sub}}}
\sum\limits_{\nu,\nu^{\prime}=1}^{N_{\textrm{sub}}}
|T_{\boldq_{\textrm{L}},\boldq_{\textrm{R}}}|^{2}
\Bigl(
n[\epsilon_{\nu}(\boldq_{\textrm{L}})]-n[\epsilon_{\nu^{\prime}}(\boldq_{\textrm{R}})]
\Bigr)
P\Bigl(
\frac{1}{\epsilon_{\nu}(\boldq_{\textrm{L}})-\epsilon_{\nu^{\prime}}(\boldq_{\textrm{R}})}
\Bigr)\notag\\
&\times 
\Bigl\{
\textrm{Im}\Bigl[
P_{\nu l}^{\dagger}(\boldq_{\textrm{L}})
P_{l^{\prime}+N_{\textrm{sub}} \nu}(\boldq_{\textrm{L}})
P_{l \nu^{\prime}}(\boldq_{\textrm{R}})
P_{\nu^{\prime} l^{\prime}+N_{\textrm{sub}}}^{\dagger}(\boldq_{\textrm{R}})
\Bigr]\notag\\
&+
\textrm{Im}\Bigl[
P_{\nu+N_{\textrm{sub}} l}^{\dagger}(\boldq_{\textrm{L}})
P_{l^{\prime}+N_{\textrm{sub}} \nu+N_{\textrm{sub}}}(\boldq_{\textrm{L}})
P_{l \nu^{\prime}+N_{\textrm{sub}}}(\boldq_{\textrm{R}})
P_{\nu^{\prime}+N_{\textrm{sub}} l^{\prime}+N_{\textrm{sub}}}^{\dagger}(\boldq_{\textrm{R}})
\Bigr]
\Bigr\}\notag\\
&+
\frac{8S^{2}}{\hbar}
\sum\limits_{\boldq_{\textrm{L}},\boldq_{\textrm{R}}}
\sum\limits_{l,l^{\prime}=1}^{N_{\textrm{sub}}}
\sum\limits_{\nu,\nu^{\prime}=1}^{N_{\textrm{sub}}}
|T_{\boldq_{\textrm{L}},\boldq_{\textrm{R}}}|^{2}
\Bigl(
1+
n[\epsilon_{\nu}(\boldq_{\textrm{L}})]+n[\epsilon_{\nu^{\prime}}(\boldq_{\textrm{R}})]
\Bigr)
P\Bigl(
\frac{1}{\epsilon_{\nu}(\boldq_{\textrm{L}})+\epsilon_{\nu^{\prime}}(\boldq_{\textrm{R}})}
\Bigr)\notag\\
&\times 
\Bigl\{
\textrm{Im}\Bigl[
P_{\nu l}^{\dagger}(\boldq_{\textrm{L}})
P_{l^{\prime}+N_{\textrm{sub}} \nu}(\boldq_{\textrm{L}})
P_{l \nu^{\prime}+N_{\textrm{sub}}}(\boldq_{\textrm{R}})
P_{\nu^{\prime}+N_{\textrm{sub}} l^{\prime}+N_{\textrm{sub}}}^{\dagger}(\boldq_{\textrm{R}})
\Bigr]\notag\\
&+
\textrm{Im}\Bigl[
P_{\nu+N_{\textrm{sub}} l}^{\dagger}(\boldq_{\textrm{L}})
P_{l^{\prime}+N_{\textrm{sub}} \nu+N_{\textrm{sub}}}(\boldq_{\textrm{L}})
P_{l \nu^{\prime}}(\boldq_{\textrm{R}})
P_{\nu^{\prime} l^{\prime}+N_{\textrm{sub}}}^{\dagger}(\boldq_{\textrm{R}})
\Bigr]
\Bigr\}.\label{eq:rewrite1}
\end{align}
\end{widetext}
Fourth, 
we rewrite some of the products of the paraunitary matrices 
in Eq. (A18) 
using some general equalities of a paraunitary matrix. 
The general equalities are Eqs. (\ref{eq:Psym1}){--}(\ref{eq:Psym4}). 
Equation (\ref{eq:Psym2}) shows 
\begin{align}
P_{l\nu}(\boldq)P_{\nu l^{\prime}+N_{\textrm{sub}}}^{\dagger}(\boldq)
=P_{l \nu+N_{\textrm{sub}}}(\boldq)P_{\nu+N_{\textrm{sub}} l^{\prime}+N_{\textrm{sub}}}^{\dagger}(\boldq)
,\label{eq:Sym1}
\end{align}
and Eq. (\ref{eq:Psym3}) shows
\begin{align}
P_{\nu l}^{\dagger}(\boldq)P_{l^{\prime}+N_{\textrm{sub}} \nu}(\boldq)
=P_{\nu+N_{\textrm{sub}} l}^{\dagger}(\boldq)P_{l^{\prime}+N_{\textrm{sub}} \nu+N_{\textrm{sub}}}(\boldq).
\label{eq:Sym2}
\end{align}
Fifth, 
we substitute Eqs. (\ref{eq:Sym1}) and (\ref{eq:Sym2}) into Eq. (A18):
\begin{widetext}
\begin{align}
J_{\textrm{S}}^{(2)}
&=
-\frac{16S^{2}}{\hbar}
\sum\limits_{\boldq_{\textrm{L}},\boldq_{\textrm{R}}}
\sum\limits_{l,l^{\prime}=1}^{N_{\textrm{sub}}}
\sum\limits_{\nu,\nu^{\prime}=1}^{N_{\textrm{sub}}}
|T_{\boldq_{\textrm{L}},\boldq_{\textrm{R}}}|^{2} 
\textrm{Im}\Bigl[
P_{\nu l}^{\dagger}(\boldq_{\textrm{L}})
P_{l^{\prime}+N_{\textrm{sub}} \nu}(\boldq_{\textrm{L}})
P_{l \nu^{\prime}}(\boldq_{\textrm{R}})
P_{\nu^{\prime} l^{\prime}+N_{\textrm{sub}}}^{\dagger}(\boldq_{\textrm{R}})
\Bigr]\notag\\
&\times 
\Bigl[
P\Bigl(
\frac{n[\epsilon_{\nu}(\boldq_{\textrm{L}})]-n[\epsilon_{\nu^{\prime}}(\boldq_{\textrm{R}})]}
{\epsilon_{\nu}(\boldq_{\textrm{L}})-\epsilon_{\nu^{\prime}}(\boldq_{\textrm{R}})}
\Bigr)
-
P\Bigl(
\frac{1+n[\epsilon_{\nu}(\boldq_{\textrm{L}})]+n[\epsilon_{\nu^{\prime}}(\boldq_{\textrm{R}})]}
{\epsilon_{\nu}(\boldq_{\textrm{L}})+\epsilon_{\nu^{\prime}}(\boldq_{\textrm{R}})}
\Bigr)
\Bigr].\label{eq:rewrite1}
\end{align}
\end{widetext}
This is Eq. (\ref{eq:J_S^2-simpler}) 
because 
the Bose-Einstein distribution function fulfills 
$1+n(\epsilon)+n(\epsilon^{\prime})=n(\epsilon)-n(-\epsilon^{\prime})$.

\section{Example of the spin current generation using the magnon BEC}

We explain an example of how to generate the spin current 
using the magnon BEC. 
As an example, 
we consider a ferromagnet. 
We first assume that 
the commensurate ferromagnetic order becomes the most stable ground state 
due to the nonperturbative Hamiltonian, such as the ferromagnetic Heisenberg interaction, 
and its spin structure is given by 
$\langle \hat{S}_{\boldi}^{x}\rangle =\langle \hat{S}_{\boldi}^{y}\rangle =0$ 
and $\langle \hat{S}_{\boldi}^{z}\rangle =S$ for all $\boldi$. 
Then, 
we consider a perturbation, such as an external magnetic field or the magnetic anisotropy, 
which induces the site-dependent 
$\langle \hat{S}_{\boldi}^{x}\rangle = \Delta S_{\boldi}$; 
the emergence of such a term 
can be alternatively regarded as 
the site-dependent misalignment of ferromagnetically ordered spins. 
This perturbation should be small so as to keep $\Delta S_{\boldi}$ 
much smaller than 
$\langle \hat{S}_{\boldi}^{z}\rangle$; 
otherwise, 
the magnet cannot be regarded as a ferromagnet 
because the combination of the nonperturbative Hamiltonian 
and the perturbation may stabilize another ordered state. 
Since the operators $\hat{S}_{\boldi}^{x}$ for the commensurate ferromagnetic order 
is proportional to $(\hat{b}_{\boldi}+\hat{b}_{\boldi}^{\dagger})$, 
this perturbation leads to $\langle \hat{b}_{\boldi}\rangle \neq 0$, 
which is site-dependent. 
Thus, the perturbation induces 
the site-dependent angle of $\langle \hat{b}_{\boldj}\rangle=B_{\boldj} e^{-i\theta_{\boldj}}$, 
and then $\boldnabla \theta_{\boldj}$ generates the spin current. 
This spin current generation using the magnon BEC is essentially the same 
as the mass flow generation in a superfluid.

\end{document}